\def\be{\begin{equation}}
\def\te{\end{equation}}
\def\bea{\begin{eqnarray}}
\def\nn{\nonumber  \\}
\def\tea{\end{eqnarray}}

\documentclass[12pt]{article}
\baselinestretch
\begin{document}
\input epsf
\title{Non-Markovian Quantum Error Deterrence by Dynamical Decoupling in a General Environment}
\author{K. Shiokawa
\thanks{E-mail address: kshiok@physics.umd.edu}
and B. L. Hu
\thanks{E-mail address: hub@physics.umd.edu}
\\ {\small Department of Physics, University of Maryland,
College Park, MD 20742, USA} }
\date{\small (July 13, 2005)}
\maketitle
\begin{abstract}
A dynamical decoupling scheme for the deterrence of errors in the
non-Markovian (usually corresponding to low temperature, short
time, and strong coupling) regimes suitable for qubits
constructed out of a multilevel structure is studied. We use the
effective spin-boson model (ESBM) introduced recently [K.
Shiokawa and B. L. Hu, Phys. Rev. A70, 062106 (2004)] as a low
temperature limit of the quantum Brownian oscillator model, where
one can obtain exact solutions for a general environment with
colored noises. In our decoupling scheme a train of pairs of
strong pulses are used to evolve the interaction Hamiltonian
instantaneously. Using this scheme we show that the dynamical
decoupling method can suppress $1/f$ noise with slower and hence
more accessible pulses than previously studied, but it still
fails to decouple super-Ohmic types of environments.
\end{abstract}

\newpage
\section{Introduction}
In quantum information science, quantum coherence and entanglement
are used as essential resources for efficient information
processing. Since the interaction of a system with its environment
can destruct the coherence and destroy the entanglement, it is
regarded as  the most serious obstacle in the realistic
implementation of quantum information processing
\cite{Dec96,PazZurek99,NielsenChuang00}.

\paragraph{Fast time low temperature strong field challenges}
In our previous paper \cite{ShiokawaHu04}, we introduced an
Effective Spin-Boson Model (ESBM), obtained as the low
temperature limit of the quantum Brownian oscillator model (QBM)
\cite{QBM0,UncPriOpenS} when the system behaves effectively like
a two-level system. In the absence of external fields, the
solutions of ESBM match with the exact solutions for the
spin-boson model\cite{LCDFGZ87,Weiss99}. At a finite temperature
and in the presence of an external field, the multilevel
structure inherent in ESBM provides a quantitative measure of the
leakage from the two-level system by transitions to other levels.
Although leakage is unavoidable in many realistic qubit systems,
most theoretical models ignore it. It is either interpreted as the
difference between the theoretical prediction and the observation
or estimated phenomenologically by crude approximations such as
invoking the Fermi Golden rule which makes the Markovian
assumption which may contradict the existent conditions.

Our model based on exact solutions for QBM enables us to estimate
leakage analytically and provides an essential tool for probing
the multilevel coherent dynamics without invoking the
conventional approximations such as Born, Born-Markov,
rotating-wave approximations (RWA). It also provides a systematic
method for studying the quantum coherence of the system behavior
in the presence of arbitrary external pulses which is used in
error correction schemes. The exact decoherence and
Rabi-oscillations in Ohmic and super-Ohmic environment were
studied in \cite{ShiokawaHu04}. We emphasize that we don't invoke
the Markov approximation which is likely to be unreliable in the
presently studied qubit models operating at \textit{low
temperatures}. In order to study the \textit{short time} dynamics
under strong and fast pulses, RWA, obtained by ignoring the
counter rotating terms averaged out at long times, is also not
suitable. Our model does not depend on these approximations and
can thus serve the special yet important purpose of analyzing the
short time dynamics under strong external fields.

\paragraph{Error Correction versus Deterrence}
One of the proposed methods to remove errors from quantum
computation architecture is the use of error correcting codes,
which enable one to detect and correct errors after they have
occurred\cite{NielsenChuang00}. By encoding qubits in a redundant
way, we can detect errors from majority voting. Although this
method works irrespective of the source of errors, actual coding
requires many physical qubits including ancillas and its
implementation is not practical at the present stage of solid
state qubits. When the system-bath interaction occurs
collectively, we can encode qubits in the decoherence free
subspace in order to avoid decoherence before it occurs\cite{DFS}.

%%%%%%%%%%%%%%%%%%%%%%%%%%%%%%%%%%%%%%%%%%%%%%%%%%%%
%DD
%%%%%%%%%%%%%%%%%%%%%%%%%%%%%%%%%%%%%%%%%%%%%%%%%%%%%%%%
In this paper, we apply the ESBM to quantum error deterrence by
dynamical decoupling, in which the external pulses are used to
overpower the influence of the environment on the system. Using
the ESBM, we give a more systematic study of the decoupling method
which hopefully provides deeper insights in these processes than
previously done based on simpler models.

In the popular scheme known as ``bang-bang" (BB) control one
introduces a sequence of square pulses
\cite{Ban:98,ViolaLloyd:98,VKL:99,Duan:98,Zanardi:99,Vitali:99,Uchiyama:02,LidarWu:03,ByrdWuLidar:04}.
Most theoretical analysis of this method are directed at
pure-dephasing in the spin-boson model: the system-bath
interaction is through the Pauli matrix $\sigma_{z}$, or the
atom-field interaction under the rotating wave
approximation\cite{Vitali:99}. Similar techniques are used for
the study of suppression of collisional decoherence
\cite{SearchBerman00} and spontaneous decay \cite{ASW01} or the
modified coherent dynamics by applying sinusoidal
waves\cite{KofmanKurizki01}.

Dynamical decoupling by spin-echo pulses has been commonly used
to suppress decoherence due to magnetic field fluctuations for
ion-trap quantum teleportation \cite{IontrapQtelportation}.
Similar to the spin-echo a pair of $\pi$ pulses are used in the
bang-bang control. In contrast, our decoupling method uses a
rapid sequence of a pair of pulses with opposite sign to rotate
the oscillator coordinate in the same direction in the phase
space by mimicking the instantaneous Hamiltonian evolution. Since
our ESBM model has more realistic features than those models
previously used to study decoupling and the advantage of being
exactly solvable, it provides an ideal testing ground for the
effectiveness of dynamical decoupling in realistic settings beyond
the regimes delimited by conventional approximations.
%%%%%%%%%%%%%%%%%%%%%%%%%%%%%%%%%%%%%%%%%%%%%%%%%%%%%%%%%%%%%%%%%%%%%%%%%%%%%

It has been argued that for the decoupling to be successful one
needs to use fast pulses with pulse duration much shorter than the
inverse frequency cutoff of the bath,
\begin{equation}
\Delta t\ll 1/\Lambda_{UV}, \label{eq:BBcond}
\end{equation}\cite{ViolaLloyd:98}. In
\cite{ShiokawaLidar04}, it was shown that when dynamical
decoupling is applied to $1/f$ noise, or more generally,
$1/f^{\alpha }$ ($ \alpha >0$), the pulses can be somewhat slower
although the condition
\begin{equation}
\Delta t< \pi/\Lambda_{UV}, \label{eq:BBcond2}
\end{equation}
 still needs to be satisfied. The
decoupling was shown to be more effective when the noise
originates from a bath with $1/f$ spectrum than in the Ohmic case
due to infrared (IR) dominance of modes in the $1/f$ spectrum.

%%%%%%%%%%%%%%%%%%%%%%%%%%%%%%%%%%%%%%%%%%%%%%%%%%%
Noise with $1/f$ type of spectrum are ubiquitous and appears in a
wide variety of circumstances, for instance, river flow, sunspot
activity, neuronal spike trains, and human
coordination\cite{Bosman01}. Various interpretations are given but
so far no explanation is universal. For electronic systems, it is
often attributed to the motion of defects or impurities  or
background charge fluctuations, although different
interpretations are possible.

In \cite{NPYT02}, an experiment was performed with spin-echo-type
pulses to identify the dominant source of decoherence in
superconducting charge qubits made of Cooper-pair boxes. The
authors concluded that a dominant source of decoherence is due to
$1/f$ charge noise. In \cite{ShiokawaLidar04}, the authors argued
that similar techniques can be used to efficiently suppress the
$1/f$ noise. They also discussed the limitation and general
applicability of decoupling pulses. Similar results were found in
the study of bang-bang control against telegraph noise\cite{BSF}.
%%%%%%

On a short time scale comparable to the bath time scale,  the
details of the system-bath interaction and internal bath dynamics
are no longer negligible. Those details will emerge when
non-Markovian features of the dynamics become important. For the
QBM model, these details are captured by the bath spectral density
$I(\omega)$. For $1/f$ noise the spectral density is concentrated
in the low frequency range. As a result, the decoupling result
depends more sensitively on the lower than on the upper cutoff.
For the pulse interval close to the threshold value, decoherence
rate can increase\cite{ShiokawaLidar04} (see Eq. (4)). Similar
accentuation of the decay process can occur from infrequent
measurements, which is known as the quantum anti-Zeno
effect\cite{KofmanKurizki00}.

In Sec. 2 we discuss the effect of a quantum Brownian harmonic
oscillator (QBO) under the influence of a general environment and
an external pulse field. We use the Fock space representation for
the harmonic QBO to define an effective two level system. In Sec.
3 we introduce dynamical decoupling (DD) in this effective
spin-boson model (ESBM) and describe how to design pulses to
minimize the influence of the environment and limit the leakage
(to higher levels). We describe how well our DD scheme work in
error deterrence for different environments.  In Sec. 4 we draw
the conclusions.

%%%%%%%%%%%%%%%%%%%%%%%%%%%%%%%%%%%%%%%%%%%%
\section{Modified quantum Brownian motion due to pulses}
\label{QBMpulse}

\subsection{The model}
Our model consists of a quantum Brownian particle interacting with
a thermal bath in the presence of external pulses. We use the
units in which $k_B=\hbar=1$. The Hamiltonian for our model can be
written as
\begin{eqnarray}
   H = H_S +  H_E + H_I + H_P,
      \label{H}
\end{eqnarray}
where the dynamics of the system $S$ (with coordinate $x$ and
momentum $p$) is described by the Hamiltonian
%$H_C = H_C( {x} {p})$,
\begin{eqnarray}
    H_S = \frac{ {p}^2}{2M} + V(x).
      \label{HC}
\end{eqnarray}
The Hamiltonian of the environment is assumed to be composed of
 harmonic oscillators with natural frequencies $\omega_{n}$ and masses $m_{n}$,
\begin{eqnarray}
 H_{E} = \sum_{n=1}^{N}( \frac{ {p}_{n}^{2}}{2 m_{n}}
    + \frac{m_{n} \omega_{n}^{2} {q}_{n}^{2}}{2}).
     \label{HB}
\end{eqnarray}
where ($ {q}_1,..., {q}_N, {p}_1,..., {p}_N)$ are the coordinates
and their conjugate momenta.
The interaction between the system S and the environment E is
assumed to be bilinear,
\begin{eqnarray}
  H_{I} =  {x} \sum_{n=1}^{N} c_{n}  {q}_{n},
    \label{HI}
\end{eqnarray}
where $c_n$ is the coupling constant between the quantum Brownian
oscillator (QBO) and the $n$th environment oscillator with
coordinate $q_n$. The coupling constants enter in the spectral
density function $I(\omega)$ of the environment defined by,
\begin{eqnarray}
I(\omega) \equiv \pi \sum_{n}\frac{c_{n}^2}{2 m_n \omega_n}
\delta(\omega - \omega_n). \label{SpectralDensity}
\end{eqnarray}
We assume the spectral density has the form
\begin{equation}
I(\omega) = 2 M \gamma \omega^{\nu} e^{-\omega/\Lambda_{UV}},
\end{equation}
where $\nu=1$ is Ohmic, $\nu<1$ is sub-Ohmic, and $\nu>1$ is
supra-Ohmic. We will study Ohmic, super-Ohmic with $\nu=3$, and
sub-Ohmic with $1/f$ type noise. Since the last case is sensitive
to the infrared (IR) regime, we introduce an infrared (IR) cutoff
$\Lambda_{IR}$ for our frequency integral. The law of large
numbers tells us that a general environment consisting of a large
number of fluctuators weakly coupled to the system can also be
reduced to a bosonic environment. This description is supported
numerically\cite{PFFF:03} and has been applied successfully to
explain the effect of $1/f$ noise on superconducting charge
qubits\cite{NPYT02}.

%%%%%%%%%%%%%%%%%%%%%%%%%%%%%%%%%%%%%%%%%%%%%%%%%%%%%%%%%%%%%%%%%%%

For a linear QBO, the potential is
\begin{eqnarray}
  V(x) = \frac{M \Omega^2 x^2}{2},
\end{eqnarray}
where $\Omega$ is the natural frequency of the system oscillator.
The pulse Hamiltonian has the form
\begin{eqnarray}
H_P= H_S \sum_{n=1}^{N}\,P_{n}(t)
    \label{HP}
\end{eqnarray}
where
\begin{eqnarray*}
P_{n}(t)=\left\{
\begin{array}{lcc}
V\hspace{3mm} & \Delta t \leq t< \Delta t +\tau &~(\mbox{mod}~ 2\Delta t +2\tau)\\
-V\hspace{3mm}& -\tau \leq t < 0 &~(\mbox{mod}~ 2\Delta t +2\tau)\\
0\hspace{3mm} & {\rm otherwise}
\end{array}
\right.  \label{Pulse}
\end{eqnarray*}
Thus $\tau$ is the duration of the pulse and $\Delta t$ is the
pulse interval between the end of one pulse and the beginning of another of
opposite sign.

\subsection{The influence functional of a system acted on by pulses}
\label{app:Formulas}

Here we will make connection with prior treatments of QBM based on
the influence functional\cite{QBM0} with a phase space
representation for the Wigner function\cite{Wigner32}. First we
define the transition amplitude between the initial state $| x_0
~q_{0} \rangle$ at $t=0$ and the final state $| x ~q \rangle$ at
time $t$ to be
\begin{eqnarray}
        {K}(x,q;t \mid x_0, q_{0};0)
 \equiv \langle x ~q | U(t,0) | x_0 ~q_{0} \rangle.
          \label{KernelK}
\end{eqnarray}
where $U(t,0)=e^{- i H t}$ is a time evolution operator. In the
presence of strong pulses, i.e. $V\rightarrow \infty$ while
keeping $V \tau $ fixed, the pulse Hamiltonian commutes with the
system-environment evolution Hamiltonian. Then the total
amplitude can be sequenced in time as
\begin{eqnarray}
        {K}(x,q;t \mid x_0, q_{0};0)&=&
      \langle x ~q | U(t,t_{2N}) U_p^{\dagger} U(t_{2N},t_{2N-1})
      U_p  U(t_{2N-1},t_{2N-2})...\nn
      & & U_p^{\dagger} U(t_{2},t_{1}) U_p  U(t_{1},0)
      | x_0 ~q_{0} \rangle\\
      &=&
      \langle x ~q | U(t,t_{2N}) U_p(t_{2N},t_{2N-1})
      U(t_{2N-1},t_{2N-2})...\nn
      & & U_p(t_{2},t_{1}) U(t_{1},0)
      | x_0 ~q_{0} \rangle
          \label{Kernelpulse}
\end{eqnarray}
where $U_{p}\equiv e^{{-iH_S V\tau}}$ and
$U_p(t_{2N},t_{2N-1})=e^{{iH_S V\tau}}U(t_{2N},t_{2N-1}) e^{{-iH_S
V\tau}}$ is the modified evolution kernel due to pulses. This form
shows that the effect of a pair of pulses is to evolve the system
variable $x$ in the interaction Hamiltonian {\it instantaneously}
to $x \cos(\Omega V\tau)+ p \sin(\Omega V\tau)/\Omega $. In phase
space, this corresponds to an instantaneous rotation by an angle
$\Omega V \tau$.

%%%%%%%%%%%%%%%%%%%%%%%%%%%%%%%%%%%%%%%%%%%%%%%%%%%%%%%%%%%%
%  Master Equation: integral form
%%%%%%%%%%%%%%%%%%%%%%%%%%%%%%%%%%%%%%%%%%%%%%%%%%%%%%%%%%%
The Liouville equation for the density matrix is
\begin{eqnarray}
i \frac{\partial}{\partial t}  {\rho}(t) = \left[ H,  {\rho}(t)
\right], \label{Liouville}
\end{eqnarray}
where $\left[ ,  \right]$ is the commutator. In the coordinate
representation, the density matrix becomes
\begin{eqnarray}
 {\rho}(x,x',q,q',t)
    \equiv \langle x ~q |  {\rho}(t) | x' ~q' \rangle
\end{eqnarray}
where we have used $q \equiv \{ q_{n} \}$ to denote the
environment variables collectively. The time evolution of the
density matrix is given by
\begin{eqnarray}
   {\rho}(x,x',q,q',t) =
       \int d x_{0} d x'_{0} d q_{0} d q'_{0}
        {K}(x,q; t \mid x_0, q_{0};0)
   {\rho}(x_0, x'_0, q_{0}, q'_{0}, 0)
    {K}^{*}(x',q'; t \mid x'_0, q'_{0};0).
          \label{DMintegral}
\end{eqnarray}

For a factorized initial condition between the system and the
environment: ${\rho}(x_0, x'_0, q_{0}, q'_{0}, 0) =
   {\rho}_{S}(x_0, x'_0, 0)
  \otimes
  \rho_{B}(q_{0}, q'_{0},0)$, after tracing out the environment harmonic oscillator
variables, we obtain an equation for the reduced density matrix:
\begin{eqnarray}
   {\rho_r}(x,x',t) \equiv \int dq
{\rho}(x,x',q,q,t) =
       \int d x_{0} d x'_{0}
        {J}_r(x, x';t \mid x_0, x'_0; 0)
        {\rho}_{S}(x_0, x'_0, 0),
            \label{DMredintegral}
\end{eqnarray}
where its time evolution operator is given by
\begin{eqnarray}
  {J}_r(x, x';t \mid x_0, x'_0; 0) =
  \int d q d q_{0} d q'_{0}
   {K}(x,q;t \mid x_0, q_{0};0)
   \rho_{B}(q_{0}, q'_{0} ,0)
    {K}^{*}(x',q; t \mid x'_0, q'_{0};0).
   \label{Jr}
\end{eqnarray}
The total action of the system ${\cal S}[x,x']$ enters as
\begin{eqnarray} J_{r}(x, x';t \mid x_0, x'_0; 0)
  \equiv \int^{(x x')}_{(x_0 x'_0)} {\mathcal D} x {\mathcal D} x'
  e^{i {\cal S}[x,x']}
\nonumber \\
\label{JF}
\end{eqnarray}
and consists of two contributions:
\begin{eqnarray}
{\cal S}[x,x']&=& {\cal S}_S[x,x']+{\cal S}_{IF}[x,x'].
\end{eqnarray}
The action for the system ${\cal S}_{S}$ is given by
\begin{eqnarray}
&&{\cal S}_{S}= \int_{0}^{t} ds \{ M\dot{R}(s)\dot{r}(s) -
M\Omega^2 R(s)r(s) \},
\end{eqnarray}
where $R \equiv (x+x')/2$, $r \equiv x-x'$.
%%%%%%%%%%%%%%%%%%%%%%%%%%%%%%%%%%%%%%%%%
The influence action ${\cal S}_{IF}[x,x']$ accounts for the effect
of the environment on the system $S$ and is given by
\begin{eqnarray}
&&{\cal S}_{IF}[R,r] = i \int_{0}^{t} ds \int_{0}^{s} ds' y(s)
\nu(s-s') y(s') \nonumber \\
&&\qquad \qquad - 2 \int_{0}^{t} ds \int_{0}^{s} ds' y(s)
\mu(s-s') Y(s'), \label{SIF}
\end{eqnarray}
where $Y(t) \equiv (x(t)+x'(t))\cos\theta(t)/2+
M(\dot{x}(t)+\dot{x}'(t)) \sin\theta(t)/2\Omega$, $y(t) \equiv
(x(t)-x'(t))\cos\theta(t)+ M(\dot{x}(t)-\dot{x}'(t))
\sin\theta(t)\Omega$. The time dependent angle $\theta(t)$ is
defined as
\begin{eqnarray}
\theta(t)=\left\{
\begin{array}{lcc}
0 \hspace{3mm} & \Delta t \leq t< 2\Delta t &~(\mbox{mod}~ 2\Delta t)\\
\Omega V \tau   \hspace{3mm}& -\Delta t \leq t <
0&~(\mbox{mod}~2\Delta t)
\end{array}
\right.  \label{angle}
\end{eqnarray}
and
\begin{eqnarray}
\mu(t)&=&-\frac{1}{\pi} \int_{0}^{\infty} d \omega I(\omega)
\sin \omega t \\
\nu(t)&=& \frac{1}{\pi} \int_{0}^{\infty} d \omega I(\omega) \coth
\frac{\beta \omega}{2} \cos \omega t,  \label{IFkernel}
\end{eqnarray}
are the dissipation and noise kernels respectively for an
environment initially at thermal equilibrium.
%%%%%%%%%%%%%%%%%%%%%%%%%%%%%%%%%%%%
We are particularly interested in the case in which each pair of
pulses rotate the system by an angle $\pi$ in the phase space,
i.e., $\Omega V \tau =(2n+1)\pi$ with an integer $n$. In this
special case, due to the periodic nature of the harmonic motion,
difference between our method and the bang-bang control is
reduced to the phase factor. The effect of these pulses is to
modify the form of the influence kernels in the action to
\begin{eqnarray}
&&{\cal S}_{IF}[R,r] = i \int_{0}^{t} ds \int_{0}^{s} ds' r(s)
\tilde{\nu}(s,s') r(s') \nonumber \\
&&\qquad \qquad - 2 \int_{0}^{t} ds \int_{0}^{s} ds' r(s)
\tilde{\mu}(s,s') R(s'), \label{SIFbb}
\end{eqnarray}
where the modified kernels $\tilde{\mu}$ and $\tilde{\nu}$ are
defined as
\begin{eqnarray}
\tilde{\mu}(s,s')&=& \cos\theta(s) \mu(s-s')
\cos\theta(s') \\
\tilde{\nu}(s,s')&=& \cos\theta(s) \nu(s-s') \cos\theta(s')
\label{modifiedkernel}
\end{eqnarray}

>From Eqs.~(\ref{SIFbb}) the Euler-Lagrange equations for $R$ and
$r$ are
\begin{eqnarray}
 M \ddot{R}(t)
+  M \Omega^2 R(t) &+& 2 \int_{0}^{t} ds \tilde{\mu}(t-s) R(s)=0,
\label{EL1}
\end{eqnarray}
\begin{eqnarray}
M \ddot{r}(t) +
 M \Omega^2 r(t)
&-& 2 \int_{s}^{t} ds' \tilde{\mu}(t-s) r(s)=0. \label{EL2}
\end{eqnarray}
The effect of the environment is included in the nonlocal kernels
$\mu$ and $\nu$. On the other hand, these kernels contain the
information of the past history of the environment as modified by
the presence of the system variables. Thus these equations respect
the self-consistency in the dynamics between the system and the
environment.

If we formally write the two independent solutions of the
homogeneous parts of Eq.~(\ref{EL1}) and Eq.~(\ref{EL2}) to be
$u_i(s)$ and $v_i(s)$, $i=1,2$, with boundary conditions
$u_1(0)=1,u_1(t)=0$, $u_2(0)=0,u_2(t)=1$ and $v_1(0)=1,v_1(t)=0$,
$v_2(0)=0,v_2(t)=1$, respectively, the solutions of these
uncoupled equations can be specified uniquely by the initial and
final conditions ($R_0, R_t$) and ($r_0, r_t$) as
\begin{eqnarray}
R_c(s) &=& R_0 u_1(s)  + R_t u_2(s),
\nonumber \\
r_c(s) &=& r_0 v_1(s)  + r_t v_2(s). \label{SolutionRr}
\end{eqnarray}
An evaluation of the path integral can be carried out which is
dominated by the classical solutions in (\ref{SolutionRr}). With
these classical solutions, we write the action ${\cal S}[x,x']$ as
\begin{eqnarray}
{\cal S}[R_c,r_c]
&=& \Big( M \dot{u}_1(t) R_0 + M \dot{u}_2(t) R_t\Big) r_t \nonumber \\
&-& \Big( M \dot{u}_1(0) R_0 +  M \dot{u}_2(0) R_t\Big)r_0 \nonumber \\
&+& i \Big( a_{11}(t) r_0^2 + ( a_{12}(t) + a_{21}(t)) r_0 r_t +
a_{22}(t) r_t^2 \Big) \;. \label{SQBM}
\end{eqnarray}
where
\begin{eqnarray}
a_{kl}(t) &=& \frac{1}{2} \int_{0}^{t} ds \int_{0}^{t} ds' v_k(s)
\tilde{\mu}(s-s') v_l(s') \label{aij}
\end{eqnarray}
(for $k,l=1,2$) contains the effects on the system dynamics from
fluctuations in the environment modified by the pulses.

%%%%%%%%%%%%%%%%%%%%%%%%%%%%%%%%%%%%%%%%%%%%%%%%%%%%%%%%%%%%%%%

Using the results above, $J_r$ in Eq.~(\ref{JF}) can be written in
a compact form,
\begin{eqnarray}
J_r(R_t, r_t;t  \mid R_0, r_0; 0)   =
     N(t) e^{i {\bf R}^{T} {\bf u} {\bf r} - {\bf
r}^{T}{\bf a} {\bf r} },
  \label{Jrsimple}
\end{eqnarray}
where $({\bf a})_{ij} = a_{ij}$, ${\bf R}^T=(R_0, R_t)$ and ${\bf
r}^T=(r_0, r_t)$
\begin{eqnarray}
{\bf u} =
        \left( \begin{array}{cc}
       u_{11}  & u_{12}   \\
       u_{21}  & u_{22}
         \end{array}      \right)
 \equiv
       M  \left( \begin{array}{cc}
       - \dot{u}_{1}(0) & \dot{u}_{1}(t)  \\
       - \dot{u}_{2}(0) & \dot{u}_{2}(t)
         \end{array}      \right)\; .
\label{U}
\end{eqnarray}

\subsection{QBM in the phase space representation}
\label{sec:PHASE}

Formally the Wigner distribution function is related to the
density matrix by
\begin{eqnarray}
W(R,P,t)= \frac{1}{2\pi} \int dr e^{-iPr} \rho_r(R+r/2,R-r/2,t).
\end{eqnarray}
It obeys the evolution equation
\begin{eqnarray}
W(R_t,P_t,t)=
 \int dR_0 dP_0 \; K(R_t, P_t;t \mid R_0, P_0; 0)
 W(R_0,P_0,0), \label{WrdensityC}
\end{eqnarray}
where $K(R, P;t \mid R_0,P_0;0)$ is the propagator for the Wigner
function defined by
\begin{eqnarray}
K(R, P;t \mid R_0,P_0;0)= \frac{1}{2 \pi} \int dr dr_0\;
e^{-i(Pr-P_0r_0)} J_r(R, r;t  \mid R_0, r_0; 0)\;.
\end{eqnarray}
It is given by
\begin{eqnarray}
 K(R, P; t \mid R_0, P_0; 0)
&=& \frac{N(t)}{2 \pi} \int dr dr_0 \;
e^{i (-P r + P_0 r_0+ {\mathcal L})} \nonumber \\
 &=& N_W(t)
%e^{ - \delta \vec{X}^{T} {\bf \Sigma}^{-1} \delta \vec{X}},
 \exp\Big[ - \delta \vec{X}^{T}
{\bf \Sigma}^{-1} \delta \vec{X} \Big],\nonumber
  \label{Ksigma}
\end{eqnarray}
where $N_W(t)=N(t)/2 \sqrt{|{\bf a}|}$ and $|{\bf a}|$ is the
determinant of ${\bf a}$. The vector $\delta \vec{X}= \vec{X}
-\langle \vec{X} \rangle$, with
\begin{eqnarray}
\vec{X} = \left( \begin{array}{c}
        R \\
        P       \end{array}      \right) \;,
\label{RWvec}
\end{eqnarray}
and
\begin{eqnarray}
\langle \vec{X} \rangle= \left( \begin{array}{c}
        \langle R \rangle \\
        \langle P \rangle
         \end{array}      \right)
=  \left( \begin{array}{cc}
       C_{11}  & C_{21}\\
       C_{12}  & C_{22}
         \end{array}      \right)
\left( \begin{array}{c}
        R_0 \\
        P_0
         \end{array}      \right)
=\frac{-1}{u_{21}}
   \left( \begin{array}{cc}
       u_{11} & 1 \\
       |{\bf u}|  & u_{22}
         \end{array}      \right)
\left( \begin{array}{c}
        R_0 \\
        P_0
         \end{array}      \right).
\label{RC}
\end{eqnarray}
Here ${\bf \Sigma}$ is a matrix characterizing the induced
fluctuations in the system from the influence of the environment:
\begin{eqnarray}
&&{\bf \Sigma} = \frac{2}{u_{21}^2}
       \left( \begin{array}{cc}
        a_{11}                       & a_{12} u_{21} - a_{11} u_{22} \\
       a_{12} u_{21} - a_{11} u_{22} &
       a_{11} u_{22}^2 - 2 a_{12} u_{21} u_{22} + a_{22} u_{21}^2
         \end{array}      \right).
\label{QW2}
\end{eqnarray}

Although many equivalent phase space representations are
known\cite{QOtext}, for our purpose, it is convenient to define
the characteristic functions obtained by the Fourier transform of
the phase space distribution functions as follows:
\begin{eqnarray}
  \chi_Q(z,\bar{z},t) =  \int d^2 \alpha Q(\alpha,\bar{\alpha})
  e^{i \bar{z}  \bar{\alpha}} e^{i z \alpha }\\
\chi_W(z,\bar{z},t) =  \int d^2 \alpha W(\alpha,\bar{\alpha})
  e^{i \bar{z}  \bar{\alpha}} e^{i z \alpha },
                \label{chiQchiWQW}
\end{eqnarray}
where $\alpha\equiv \sqrt{\Omega/2} R+i\sqrt{1/2\Omega}P$. The
characteristic function $\chi_Q(z,\bar{z})$ for the Q-distribution
is related to $\chi_W(z,\bar{z})$, the characteristic function
for the Wigner distribution, as
\begin{eqnarray}
  \chi_Q(z,\bar{z},t) =  e^{- \frac{|z|^2}{2}}
  \chi_W(z,\bar{z},t).
                \label{chiQchiW}
\end{eqnarray}
For instance, the characteristic function $\chi_Q(z,\bar{z},t)$
for the harmonic QBO evolved from the initial ground state has the
following Gaussian form:
\begin{eqnarray}
 \chi_Q(z,\bar{z},t) = \exp\left[
 - \frac{\langle a^2(t) \rangle}{2} z^2
 - \frac{\langle a^{\dagger 2}(t) \rangle}{2} \bar{z}^2
 -\langle a(t)a^{\dagger}(t) \rangle|z|^2
\right]. \label{chiQgaussian}
\end{eqnarray}
The time dependent coefficients are anti-normal ordered operator
averages of second moments given by $\langle a^2\rangle=c+\sigma$,
$\langle a a^{\dagger}\rangle=C+\Sigma+1/4$, which have their
origins in the classical trajectory ${\bf C}$ of a damped harmonic
oscillator in Eq. (\ref{RC}), the induced fluctuations ${\bf
\Sigma}$ from the environment in Eq.(\ref{QW2}). The relations of
these components can be written as follows:
\begin{eqnarray}
8 c&=&  C_{22}^2 -  C_{11}^2 +  \Omega^2 C_{12}^2
-\frac{C_{21}^2}{\Omega^2} + 2 i (C_{11}C_{12}+ C_{21}C_{22}) \nn
8 C&=& C_{11}^2 + \Omega^2 C_{12}^2 + \frac{C_{21}^2}{\Omega^2} +
C_{22}^2 \nn
 4 \sigma&=&    \Omega \Sigma_{11} -
\frac{\Sigma_{22}}{\Omega} +2 i \Sigma_{12} \nn 4 \Sigma&=& \Omega
\Sigma_{11} + \frac{\Sigma_{22}}{\Omega}\label{sS}
\end{eqnarray}

% moved old Sec. 3.3 to here %%%%%%%%%%%%%%%%%%%%%%%%%%%%%%%%%%%%%%%%%%%%%%%%%%%%%%%%%%%%

\subsection{Fock states/phase space correspondence}
\label{sec:Fockphase}

The harmonic QBM
 \begin{eqnarray}
    H_S^{(\infty)} +H_{I}^{(\infty)} = \Omega a^{\dagger} a + \sqrt{2 \Omega}
    (a + a^{\dagger}) \sum_{n=1}^{N_B} c_{n}
    {q}_{n}
\end{eqnarray}
can be viewed as a limiting case of a finite $N$-level system:
\begin{eqnarray}
H_S^{(N)} +H_{I}^{(N)} = \Omega S_{N}^{+} S_{N}^{-} + (S_{N}^{-} +
S_{N}^{+}) \Gamma_B
         \label{NLS}
\end{eqnarray}
where $N\rightarrow \infty$. Here $\Gamma_B\equiv\sum_{n=1}^{N_B}
\tilde{c}_{n} {q}_{n}$ and we have absorbed the factor
$\sqrt{2\Omega}$ by defining $\tilde{c}_{n}= \sqrt{2 \Omega}
c_{n}$. $S_{N}^{-}$ is a number lowering operator whose matrix
element is $(S_{N}^{-})_{i j}\equiv \sqrt{i} \delta_{i+1,j}$ and
its conjugate $S_{N}^{+}$ is a number raising operator.

The correspondence  between the Fock state representation for the
pseudo spin qubits and the phase space representation of a
density matrix is given by
\begin{eqnarray}
 \rho_{kl}(t) &=&
  \int \frac{d^2 z}{\pi}\chi_Q(z,\bar{z},t)
  \langle k |  e^{- i \bar{z} a^{\dagger} } e^{- i z a } | l
  \rangle.
                \label{FockPhase}
\end{eqnarray}
>From Eq. (\ref{FockPhase}) we can directly evaluate the density
matrix in the Fock representation for an arbitrary quantum
number. The Pauli spin representation from the lowest two level
Fock representation are obtained as
\begin{eqnarray}
 \langle \sigma_x(t)\rangle &=& {\rho}_{01}(t) + {\rho}_{10}(t)\nn
 \langle \sigma_y(t)\rangle &=& i {\rho}_{10}(t) - i{\rho}_{01}(t)\nn
 \langle \sigma_z(t)\rangle &=& {\rho}_{11}(t) - {\rho}_{00}(t).
             \label{SPRep}
\end{eqnarray}
In \cite{ShiokawaHu04}, we explained in some detail the level
reduction mechanism and the relation between ESBM and other
models of quantum computing. Various models used in the study of
quantum computation can be obtained from our ESBM in different
limits.

At a finite temperature $T$, only  modes in the harmonic
oscillator up to $N \sim k_B T / \hbar \Omega$ are excited. At
sufficiently low temperature $ k_B T \sim \hbar \Omega$, the
effective excitable number of levels of a harmonic QBO is
significantly reduced. In particular, at $k_B T < \hbar \Omega$,
we expect that the system is effectively reduced to two-levels.
In this limit, our system and interaction Hamiltonians are
reduced to those of the spin-boson model
\begin{eqnarray}
H_S^{(2)} +H_{I}^{(2)} = \Omega (S_{2}^{z}+\frac{1}{2}) +
S_{2}^{x} \Gamma_B.
      \label{TLR}
\end{eqnarray}
Simplified model of decoherence with a pure dephasing
term\cite{Unruh95} can be obtained as an adiabatic limit of ESBM
when the Brownian oscillator is frozen, i.e. $\Omega \rightarrow
0$ with an unitary change of basis. This model is suitable for the
study of short time dynamics but it overestimates the decoherence
as we will see in the next section.  Dissipative system-bath
coupling without counter rotating terms common in many quantum
optics texts\cite{QOtext} is obtained by imposing a rotating wave
approximation (RWA) on ESBM.
\begin{eqnarray}
  H_S +H_{SB}
 = \Omega S_{2}^{+} S_{2}^{-} + (S_{2}^{-} + S_{2}^{+}) \Gamma_B
  \rightarrow
   H_{S} +H_{RWA} = \Omega S_{2}^{+} S_{2}^{-} + \sum_{n=1}^{N_B}
   \tilde{c}_n ( S_{2}^{-} {b}_{n}^{\dagger}
    + S_{2}^{+} {b}_{n} ),
      \label{RWA}
\end{eqnarray}
where ${b}_{n}= (\omega_n q_n +i p_n)/\sqrt{2 \omega_n}$ are bath
annihilation operators. The dynamics under this approximation
cannot capture the fast dynamics at time scales less than the
natural time scale of the system. Thus it is inadequate in the
study of generic environments.

More generally, one can split the components of $S_{N}^{+}
S_{N}^{-}$ as $S_{2}^{+} S_{2}^{-}+S_{N-2}^{+} S_{N-2}^{-}$,
where $S_{N-2}^{\pm}$ are raising and lowering operators acting on
the space outside of the lowest two levels. Similarly,
$S_{N}^{+}+S_{N}^{-} =S_{2}^{+}+S_{2}^{-}+S_{N-2}^{+}+S_{N-2}^{-}
=S_{2}^{x}+S_{N-2}^{x}+T_{2,N-2} $, where $S_{N-2}^{x}$ acts only
on the space outside of the lowest two levels and
$(T_{2,N-2})_{nm}=\sqrt{2} \delta_{n2} \delta_{m3}$ mixes the two
sectors (inside and outside of the lowest two levels) and is thus
responsible for the leakage.

%Only under such conditions
%will the formal replacement of the harmonic oscillator
%annihilation/creation operators $a,a^{\dagger}$ by the two-level
%pseudo spin annihilation/creation (Pauli) operators
%$S_{2}^{-},S_{2}^{+}$ be justifiable.

%%%%%%%%%%%%%%%%%%%%%%%%%%%%%%%%%%%%%%%%%%%%%%%%%%%%%%%%%%%%%%
\section{Dynamical Decoupling}
\label{BBESBM}

\subsection{General Scheme}

Quantum error correction based on coding schemes requires many
physical qubits to encode information\cite{NielsenChuang00}. In
more realistic situations where only a few physical qubits are
available, the implementation of quantum information processing
(QIP) requires more practical methods to prevent decoherence. This
is similar to the refocusing techniques frequently used in NMR,
where a sequence of pulses is used to modify the influence of
environment on the system. In the bang-bang control scheme, a
rapid sequence of a pair of $\pi$ pulses is used to reduce the
effect of decoherence.  In the scheme described in Sec. 2, a train
of a pair of pulses with opposite signs to change the influence of
the environment on the system.

\paragraph{Pulses to diminish environmental influence} In order to successfully
diminish the effect of the environment on the system   we require
the pulses to satisfy the following requirements:  First, the
pulses are assumed to be strong enough so that we can ignore $H_S$
during the pulse operation compared to $H_P$. In this strong
pulse limit while keeping $\Omega V \tau = (2l+1) \pi
~(l=0,1,2...)$ fixed, we obtain $e^{i H_p \tau} H_I e^{-i H_p
\tau}= -H_I $. Second, the pulses are assumed to be acting at
sufficiently short intervals $\Delta t$. Once these conditions are
satisfied, the repeated application of pulses will change the
whole system evolution to
\begin{eqnarray}
U(t=2{\cal N}\Delta t)&=& \left[ e^{i H_p \tau} e^{-i \Delta t
(H_S+H_E+H_I)}
       e^{-i H_p \tau} e^{-i \Delta t (H_S+H_E+H_I)}
\right]^{\cal N}\nn &=& e^{-i \Delta t (H_S+H_E-H_I)} e^{-i
\Delta t (H_S+H_E+H_I)}...\nn & & e^{-i \Delta t
(H_S+H_E-H_I)}e^{-i \Delta t (H_S+H_E+H_I)}    \nn &\rightarrow&
e^{-i t (H_S+H_E)}
\end{eqnarray}
where the limit with vanishing pulse intervals is taken: $\Delta t
\rightarrow 0, {\cal N}\rightarrow \infty$, while $t=2{\cal
N}\Delta t$ is kept fixed. Thus under these conditions, the
effect of the environment can be successfully eliminated.

Note that the complete elimination of the environment is achieved
only in the idealized limit $\Delta t \rightarrow 0$. In
practice, it seems sufficient to add pulses faster than the
environment characteristic scale such that the environment is
effectively frozen in the duration of the pulses. It has been
argued that this condition requires $\Delta
t<<1/\Lambda_{UV}$\cite{ViolaLloyd:98}, which is already too
stringent for the known practical implementations. Thus it is
worth examining whether this condition $\Delta t\ll
1/\Lambda_{UV}$ is strictly necessary. It is also important to
see from our pulse Hamiltonian (\ref{Pulse}) that the total
energy will be unchanged from our pulse operations.

\paragraph{Pulses to minimize leakage} In a multilevel system if we are only
interested in limiting the leakage (see previous section) from the
two lowest levels to the higher levels, we can shape the pulses to
realize the decoupling operator\cite{ByrdWuLidar:04}
\begin{eqnarray}
U_D = e^{i \phi} \left( \begin{array}{cc}
       -I_{(N-2) \times (N-2)}  & 0   \\
        0  & I_{2 \times 2}
         \end{array}      \right),
\end{eqnarray}
where $\phi$ is a phase factor. The condition $\{ U_D,
T_{2,N-2}\}=0$ together with $\left[ U_D, H_S\right]=\left[ U_D,
H_E\right]=\left[ U_D, S_{2}^{x}\right] =\left[ U_D,S_{N-2}^{x}
\right]=0$ yields
\begin{eqnarray}
U(t=2{\cal N}\Delta t) &=& \left[ U_D^{\dagger} e^{-i \Delta t
\left\{H_S+H_E+(S_{2}^{x}+S_{N-2}^{x}+T_{2,N-2})\Gamma_B\right\}}
U_D e^{-i \Delta t
\left\{H_S+H_E+(S_{2}^{x}+S_{N-2}^{x}+T_{2,N-2})\Gamma_B\right\}}
\right]^{\cal N}\nn &=& e^{-i \Delta t
\left\{H_S+H_E+(S_{2}^{x}+S_{N-2}^{x}-T_{2,N-2})\Gamma_B\right\}}
    e^{-i \Delta t \left\{H_S+H_E+(S_{2}^{x}+S_{N-2}^{x}+T_{2,N-2})\Gamma_B\right\}}...\nn
& & e^{-i \Delta t
\left\{H_S+H_E+(S_{2}^{x}+S_{N-2}^{x}-T_{2,N-2})\Gamma_B\right\}}
    e^{-i \Delta t \left\{H_S+H_E+(S_{2}^{x}+S_{N-2}^{x}+T_{2,N-2})\Gamma_B\right\}}\nn&\rightarrow&  e^{-i t \left\{H_S+H_E+(S_{2}^{x}+S_{N-2}^{x})\Gamma_B\right\}}
\end{eqnarray}
where the limit with vanishing pulse intervals is taken: $\Delta t
\rightarrow 0, {\cal N}\rightarrow \infty$, while $t=2{\cal
N}\Delta t$ is kept fixed. The leakage can be eliminated in this
limit. This method, however, will not change the coherence
property within the computational Hilbert space. Thus decoherence
within such space is not preventable with this method. We now
discuss the dynamical decoupling scheme which protects the
computational Hilbert space from the influence of the environment.

%%%%%%%%%%%%%%%%%%%%%%%%%%%%%%%%%%%%%%%%%%%%%%%%%%%%%%%%%%%%%%%%%%%%%%%

\subsection{Dynamical Decoupling in QBM:
Quantum Control of the Uncertainty Relation} \label{sec:QBMBB}

We detour a bit here to show how dynamical decoupling can alter
the uncertainty relation in quantum mechanics. The uncertainty
principle \cite{Heisenberg27} is a basic requirement for any
quantum states to satisfy. In an open quantum system, information
in the system is partly lost to the environment,  resulting in a
greater uncertainty in the quantum states of the open system than
in the original closed system. The modified uncertainty relation
in QBM in Ohmic environment has been studied
before\cite{UncPriOpenS}. Here we study the uncertainty relation
in general environment. A suitable measure for the uncertainty of
an open quantum system is
\begin{eqnarray}
A^2 &\equiv& (\Delta R )^2 (\Delta P )^2 -\frac{1}{4} \langle
\{\Delta R,\Delta P \}\rangle^2.
  \label{A2}
\end{eqnarray}
The uncertainty principle for open systems requires that for any
physical state (pure or mixed, equilibrium or nonequilibrium)
should satisfy $A \geq 1/2$.

Using the identity
\begin{eqnarray}
\langle a a^{\dagger} \rangle^2 - \langle a^2 \rangle \langle
a^{\dagger 2} \rangle = A^2 + \frac{\langle R^2 \rangle}{2}+
\frac{\langle P^2 \rangle}{2} +\frac{1}{4}
  \label{AtoRP}
\end{eqnarray}
and $\langle R^2 \rangle + \langle P^2 \rangle \geq 1$ from the
usual uncertainty principle, we obtain $\langle a a^{\dagger}
\rangle^2 - \langle a^2 \rangle \langle a^{\dagger 2} \rangle
\geq 1$.
Combining this with (\ref{chiQgaussian}) and (\ref{FockPhase}) yields
the ground state population $\rho_{00}=(\langle a a^{\dagger}
\rangle^2 - \langle a^2 \rangle \langle a^{\dagger 2} \rangle)^{-1/2}\leq 1$.
This can also be inferred directly from a probability requirement.
From our Fock state/phase space correspondence in
(\ref{FockPhase}), we can reconstruct the temporal phase space
uncertainty relation arbitrarily far from equilibrium by
observing a relaxation process in Fock space.
In Fig. 1 (a), the time evolution of the uncertainty function $A$
is plotted. We set $M=1$ for all figures. For a state initially
in a pure state, the uncertainty at $t>0$ is always greater than
the initial value.

Since decoupling pulses will reduce or
eliminate system-environment interaction and suppress the
information loss in the system, the uncertainty can be used as a
measure of the suppression of the environment-induced effect. In
particular, after the application of sufficiently strong and
frequent pulses, we expect that the states will obey an
uncertainty relation close to the initial relation, when the
environment is almost completely decoupled from the system.

In Fig. 1 (b), the uncertainty function at $t=1$ is plotted as a
function of the pulse frequency parameter $\eta \equiv
\Lambda_{UV}\Delta t/\pi$. Note that arbitrary pulses do not
always suppress the uncertainty. For slow pulses, the uncertainty
is increased rather than decreased. The peak around $\eta \sim 1$
is ascribed to the resonance between the pulse and the
environmental mode. For super-Ohmic environment, the resonance is
particularly eminent, while for $1/f$ environment, the resonance
is absent and the suppression of uncertainty is effective
throughout the whole range of the pulse frequency in the figure.
This phenomenon is related to so called quantum anti-Zeno effect.
We will discuss this issue in Sec. 3.3. For pulses faster than
the resonance $\eta<1$, the uncertainty function decreases as the
pulses become faster and for sufficiently fast pulses, the
initial uncertainty returns for all environments.

%%%%%%%%%%%%%%%%%%%%%%%%%%%%%%%%%%%%%%%%%%%%%%%%%%%%%%%%%%%%%%%%%%%%%%%
\begin{figure}[h]
 \begin{center}
\epsfxsize=.50\textwidth \epsfbox{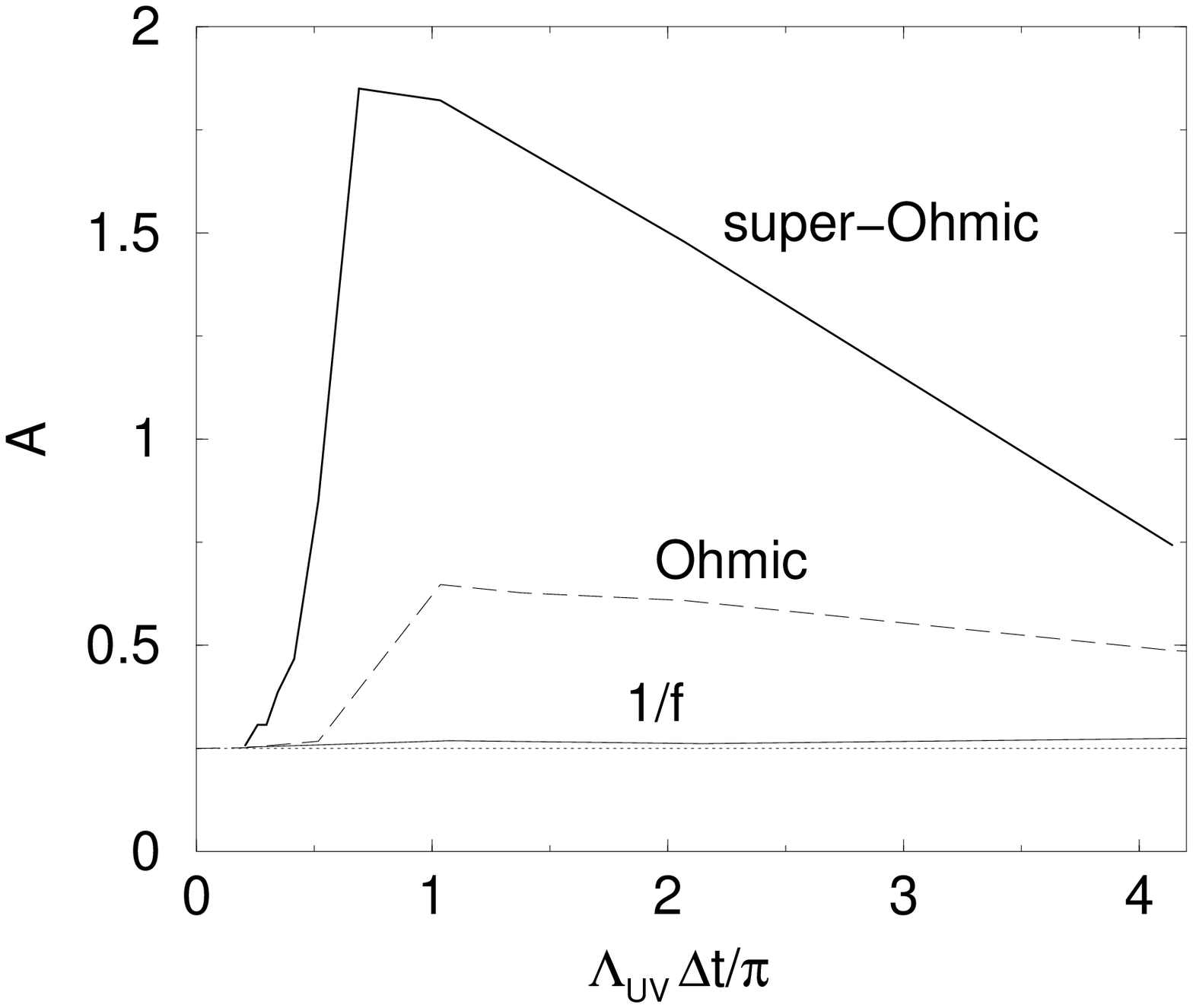}
\epsfxsize=.50\textwidth \epsfbox{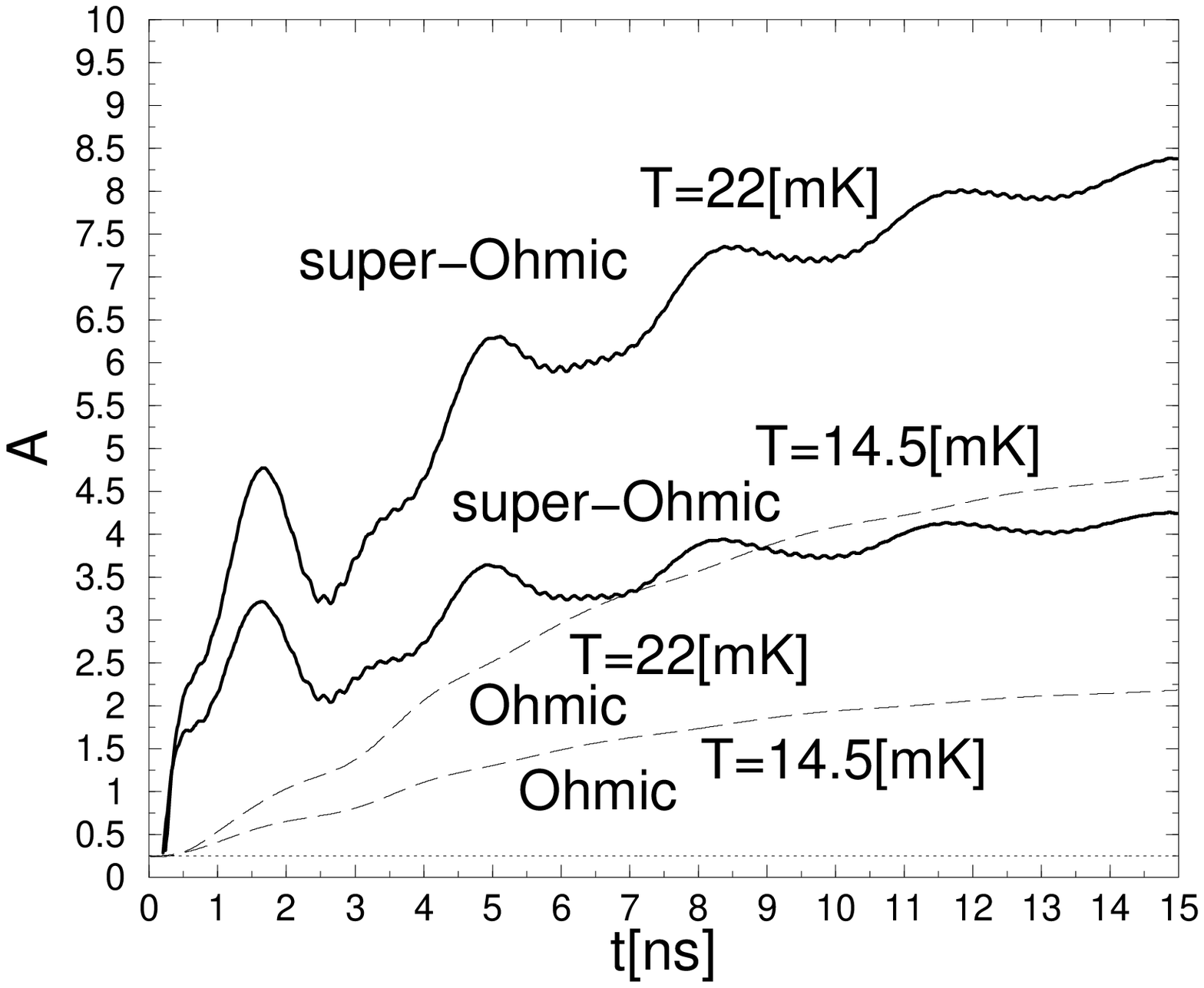}
  \end{center}
\caption{(a) The exact temporal evolution of the uncertainty
function (\ref{A2}) in units of $\hbar^2$
of a system initially in its ground state
interacting with an Ohmic and super-Ohmic environment is plotted.
%The solid curves are at $T=20[mK]$, $T=10[mK]$,
%and $T=0[mK]$ from the top of the graph down.
The dotted line indicates the minimum uncertainty at $1/4$. All
the curves start off from this value at $t=0[ns]$ and stay above
the line. The effect of vacuum fluctuations of the environment
sets in at $t\sim  0.1[ns]$  and the effect of thermal
fluctuations of the environment at $t\sim  1/T=0.3-0.5[ns]$ for
Ohmic case. Both effects set in earlier for super-Ohmic case.
%, which is $0.5[ns]$, $1[ns]$, and $\infty$ from the top down.
Other parameters are $\omega=1[GHz]$, %$\omega=0.5[GHz]$,
$\gamma=0.1[GHz]$, $\Lambda_{UV}=10[GHz]$.
(b) Same system now in the presence of
decoupling pulses at $t=0.25[ns]$.  Other parameters are
$\Omega=1[GHz]$. $T=14.5[mK]$,  $\gamma=0.1[GHz]$,
$\Lambda_{UV}=100[GHz]$ for Ohmic, $T=14.5[mK]$,
$\gamma=0.0002[GHz]$, $\Lambda_{UV}=50[GHz]$ for super-Ohmic,
$T=0[mK]$, $\gamma=0.1[GHz]$, $\Lambda_{UV}=100[GHz]$ for 1/f
environments.}
\end{figure}

\subsection{Dynamical Decoupling in ESBM}
\label{sec:ESBMDD}

In \cite{ShiokawaLidar04}, it was shown that the bang-bang pulses
with their pulse duration close to but less than the threshold
value $\pi/\Lambda_{UV}$ can eliminate $1/f$ noises efficiently.
A pure dephasing type model\cite{Unruh95} was used to illustrate
the case as follows. In the presence of the decoupling pulses, at
$t_{2{\cal N}}=2{\cal N}\Delta t$, coherence between the two
levels is given in the interaction representation by
\begin{eqnarray}
\rho _{01}^{I}(t_{2{\cal N}})=e^{-D_{P}(t_{2{\cal N}})}\,\rho _{01}^{I}(0)\;,
\label{lab2}
\end{eqnarray}
where
\begin{eqnarray}
D_{P}(t_{2{\cal N}})\sim \int_{\Lambda _{0}}^{\Lambda_{UV }}
d\omega \,I(\omega )\coth \left( \frac{\beta \omega }{2}\right)%\nn
\, {\frac{1-\cos \omega t_{2{\cal N}}}{\omega ^{2}}}\,\tan
^{2}\left( \frac{\omega \Delta t}{2}\right) \; \label{D1/f}
\end{eqnarray}
The $tan^2$ factor contains the effect of decoupling pulses. This
form indicates that decoherence will be suppressed for
\begin{eqnarray}
  \Lambda_{UV}\Delta t<\pi.
\end{eqnarray}
Indeed decoherence from all the modes with
$(4l+1)\pi/2<\omega\Delta t<(4l+3)\pi/2$ for an integer $l$ will
be enhanced. There is also a resonance of decoupling pulses when
the pulses are synchronized with the particular mode in an
environment causing a divergence in the decoherence rate that
leads to rapid destruction of coherence. Bang-bang method
suppresses $1/f$ noise for sufficiently fast pulses, but the
condition (\ref{eq:BBcond2}) still needs to be satisfied.
%%%%%%%%%%%%%%%%%%%%%%%%%%%%%%%%%%%%
% BL Changes begin here:
%%%%%%%%%%%%%%%%%%%%%%%%%%%%%%%%%%%%

At a finite temperature when thermal fluctuations begin to affect
the system dynamics, a time scale $t_{thermal}$ sets an
additional constraint for decoupling. In order for the decoupling
to be successful, the pulse interval needs to be smaller than the
smaller value of $1/\Lambda_{UV}$ or
$t_{thermal}$\cite{Vitali:99}.

Here we will show that this resonance effect does not appear in
more realistic ESBM models.
%For a generic environment, even
%at $T \sim 0$, $1/\Lambda_{UV}$ is not equivalent to the environment
%characteristic time $t_{environment}$.
$t_{environment}$ is not just a function of $\Lambda_{UV}$ but
depends on detailed characteristics of the environment. In the
case of our ESBM, $t_{environment}$ also sensitively depends on
the spectral density. In more general cases, there are many more
parameters to determine $t_{environment}$, which should be
determined by the spectroscopy of the environment.

Fig. 2 shows the time evolution of the logarithm of the decay
factor $\Gamma(t)\equiv -\log P_1$ at $T=0$ for different
environments. Sufficiently fast pulses are chosen so that the
decay rate is suppressed by the decoupling pulses.  In the
super-Ohmic case, the short time decay is remarkable due to
ultra-violet mode dominance in this environment, while in the
$1/f$ case, long time decay is drastic due to infrared mode
abundance in this environment. The efficiency of decoupling
directly reflects this characteristic. In the super-Ohmic case,
the suppression of decay rate is relatively small, while in the
$1/f$ case, improvement is rather remarkable.

In order to quantify how slow the pulses can be, let us introduce
the parameter $\eta \equiv \Lambda_{UV}\Delta t/\pi$. $\eta < 1$
means that the pulses for efficient suppression of decoherence in
this environment need to be fast compared to the UV cutoff. This
condition is difficult to be met experimentally.  $\eta > 1$ means
that the pulses can be reasonably slow and relatively easy to
implement. For 1/f noise, where $\eta=1.5>1$, our DD scheme works
very well. For Ohmic environment, $\eta =0.15<< 1$, and for
super-Ohmic environment, $\eta =0.05<< 1$, thus the pulses to
suppress Ohmic and super-Ohmic environment need to be ultra-fast.
We see that super-Ohmic environments can pose a serious obstacle
for the implementation of coherent quantum operations.

In Fig. 3 (a), coherence as a function of the pulse interval is
plotted in the pure-dephasing case. For super-Ohmic case, the
suppression is the smallest and the condition for efficient
suppression is much more  stringent than $\Lambda_{UV} \Delta t<
\pi$, while for  $1/f$ noise, the pulses with $\Delta t$ close to
$\pi/\Lambda_{UV}$ remain effective. However, irrespective of the
environment, when the pulse interval gets longer, $\Lambda_{UV}
\Delta t\sim \pi$, there is a crossover from decoherence
suppression to decoherence accentuation. There is a sharp
resonance point at $\Lambda_{UV} \Delta t= \pi$, where
decoherence rate diverges for all environment and the condition
$\Lambda_{UV} \Delta t< \pi$ need be strictly enforced.

In Fig. 3 (b), the effect of decoupling pulses in Eq. (\ref{HP})
is studied in our ESBM. The main difference from Fig. 2 (a) is
that: (i) there is no sharp resonance as seen in Fig. 3 (a) in any
environment although the crossover from decoherence suppression
to accentuation can still be seen. (ii) the condition
$\Lambda_{UV} \Delta t < \pi$ is not necessary for the
suppression of $1/f$ noise. Even very slow pulses $\Lambda_{UV}
\Delta t >> \pi$ can suppress it. The second condition (ii)
expressed before in Eq. (2) is particularly important in that it
is often too stringent for practical implementation with qubits.
Note that the crossover behavior expressed in (i) is already
manifest in the uncertainty relation plotted in Fig. 1(b).
Similar effect also appears in the study of quantum Zeno effect
where sufficiently frequent measurements suppress the population
decay\cite{MisraSudarshan77}. When the measurement is performed
less frequently, there is a crossover from quantum Zeno effect to
the so called quantum anti-Zeno effect,  where measurement
accelerates the decay\cite{KofmanKurizki00}.

%%%%%%%%%%%%%%%%%%%%%%%%%%%%%%%%%%%%%%%%%%%%%%%%%
\begin{figure}[h]
 \begin{center}
\epsfxsize=.50\textwidth \epsfbox{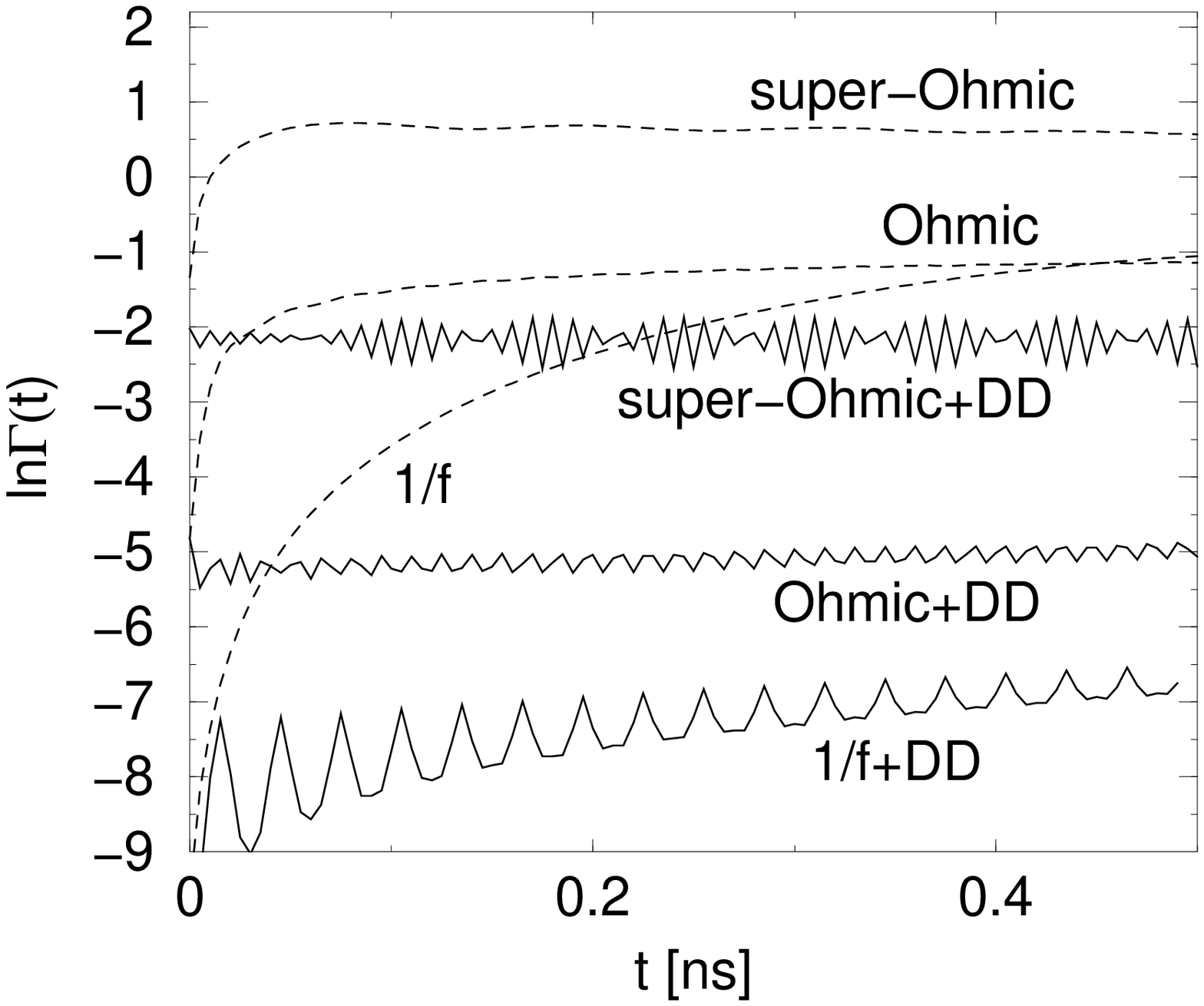}
  \end{center}
\caption{ Plots of the logarithm of the decay factor at $T=0$,
$\Omega=1[GHz]$. Solid (Dashed) curves are decay with (without)
decoupling pulses. Parameter values are: $\Lambda_{UV}=100[GHz]$,
$\gamma=0.1[GHz]$, $N=100$ for Ohmic, and $\Lambda_{UV}=100[GHz]$,
$\gamma=0.5[GHz]$, $N=30$ for 1/f, $\Lambda_{UV}=30[GHz]$, $\gamma=0.01[GHz]$,
$N=100$ for super-Ohmic case.
%von-Neumann entropy rate (thick solid line) at $T=0$ versus time.
%$\lambda_R=1$, $M_R=1$, $\omega_{\infty}=10$. Panel (a) is for the
%hmic environment with $\gamma=0.2$ and Panel (b) is for the $\nu=3$
%super-Ohmic environment with $\gamma=0.05$. The dashed line is a constant
%value with $\lambda - \gamma + O(\omega_{\infty}^{-1}) = 0.8$ for
%the Ohmic environment, $\lambda + \gamma \lambda^2
%+O(\gamma^2,\omega_{\infty}^{-1})= 1.05$ for the super-Ohmic
%environment.
}
\end{figure}

%%%%%%%%%%%%%%%%%%%%%%%%%%%%%%%%%%%%%%%%%%%%%%%%%
\begin{figure}[h]
 \begin{center}
\epsfxsize=.55\textwidth \epsfbox{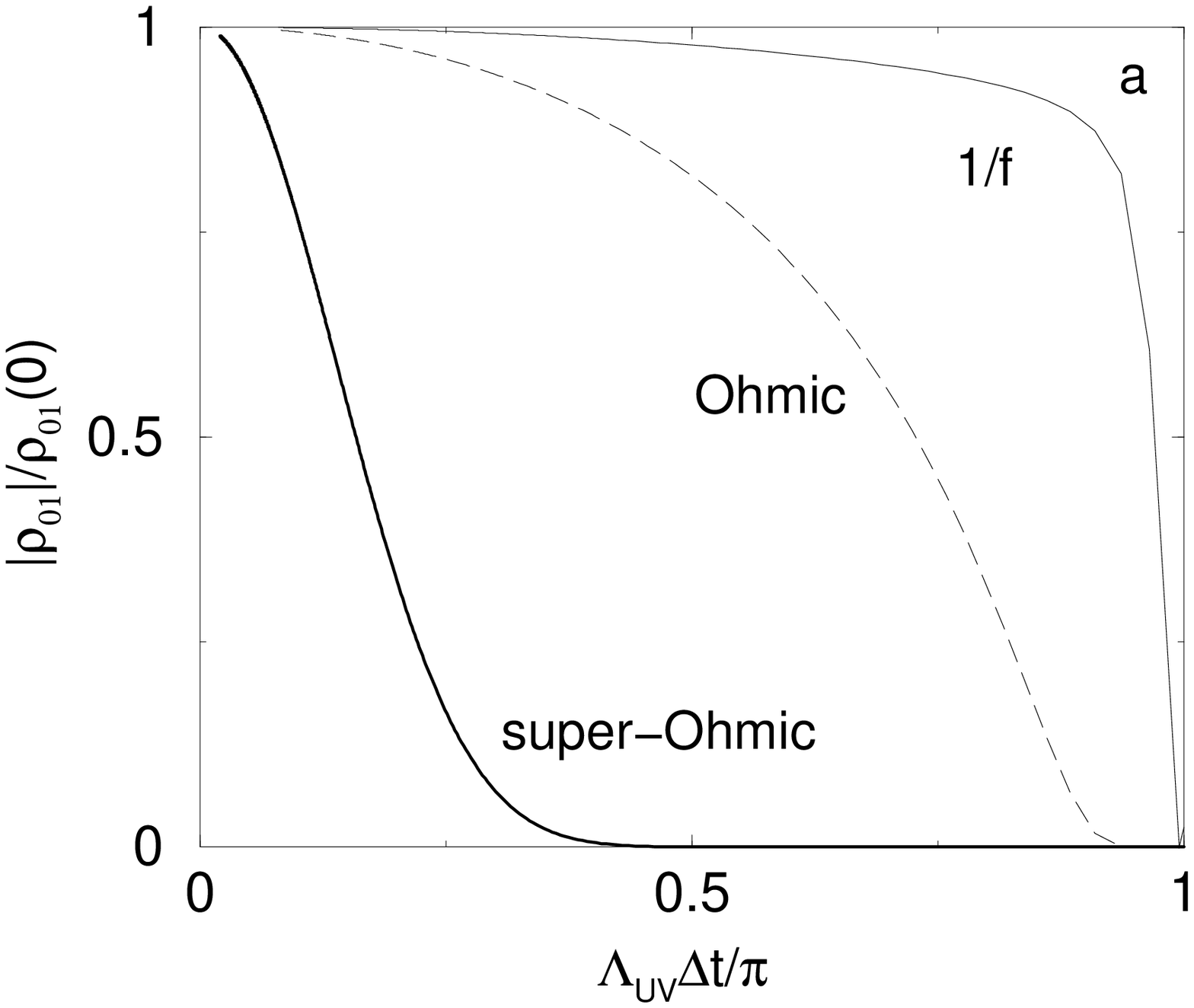}
\epsfxsize=.50\textwidth \epsfbox{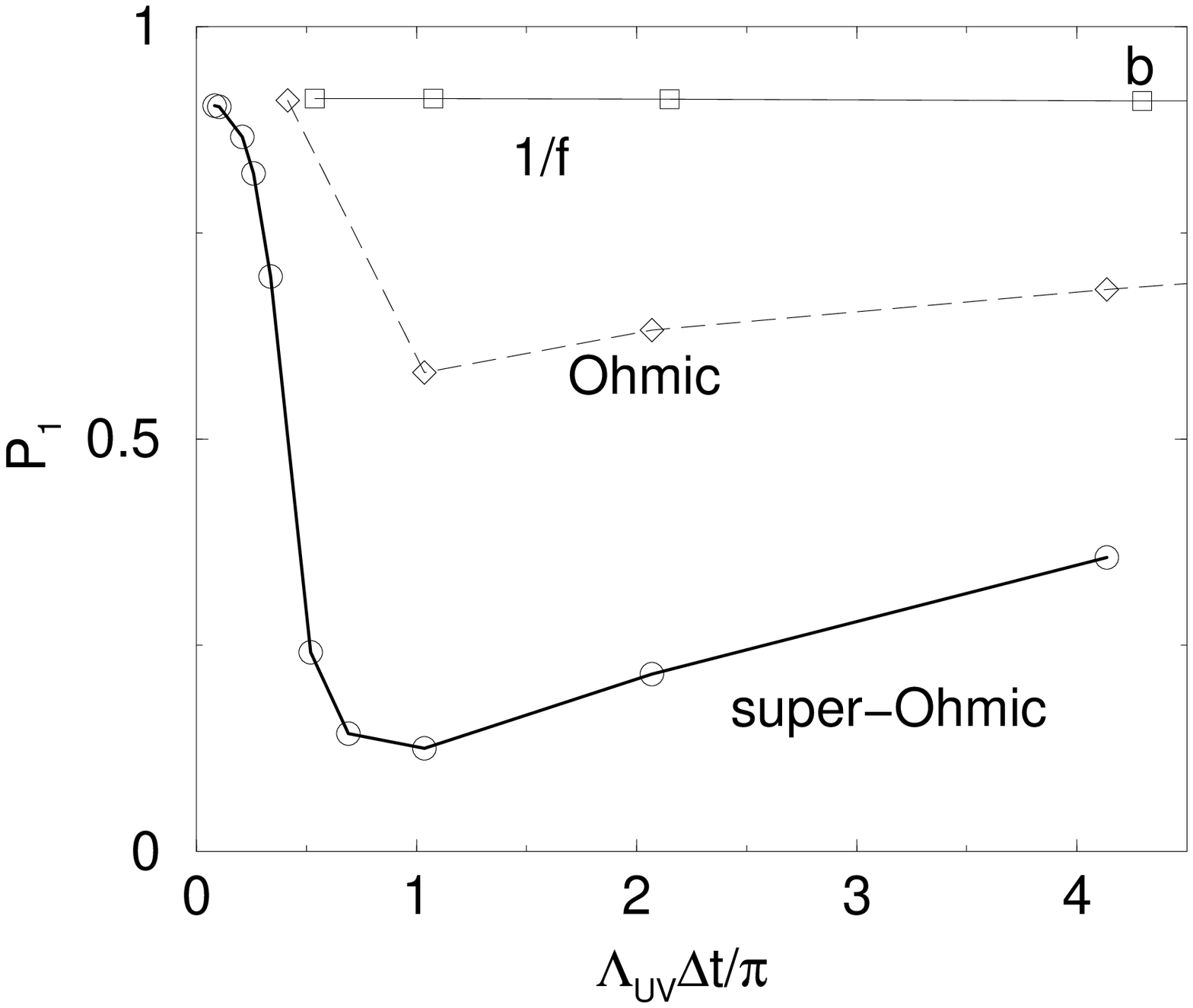}
  \end{center}
\caption{ (a) Coherence as a function of the pulse interval at
$T=0$ is plotted. $t=0.25[ns]$. Parameters are:
$\Lambda_{UV}=100[GHz]$, $\Lambda_{IR}=1$. $\gamma=0.2[GHz]$ for
$1/f$,  $\gamma=0.125[GHz]$ for Ohmic, and $\gamma=0.005[GHz]$ for
super-Ohmic case. (b) Plots of the first excited state population
at $T=0$ and $t=0.15[ns]$. Other parameter values are:
$\Lambda_{UV}=100[GHz]$, $\Omega=10[GHz]$, $\gamma=0.1[GHz]$ for
Ohmic, and $\Lambda_{UV}=100[GHz]$, $\Omega=15[GHz]$,
$\gamma=0.5[GHz]$ for 1/f, $\Lambda_{UV}=50[GHz]$,
$\Omega=15[GHz]$, $\gamma=0.01[GHz]$ for super-Ohmic case.
Parameter values are chosen to fit all the curves in the parameter
range in the figure. For the same parameter values, while their
qualitative behavior remains similar, the differences in their
values are more striking. }
\end{figure}

%%%%%%%%%%%%%%%%%%%%%%%%%%%%%%%%%%%%%%%%%%%%%%%%%
\begin{figure}[h]
 \begin{center}
\epsfxsize=.50\textwidth \epsfbox{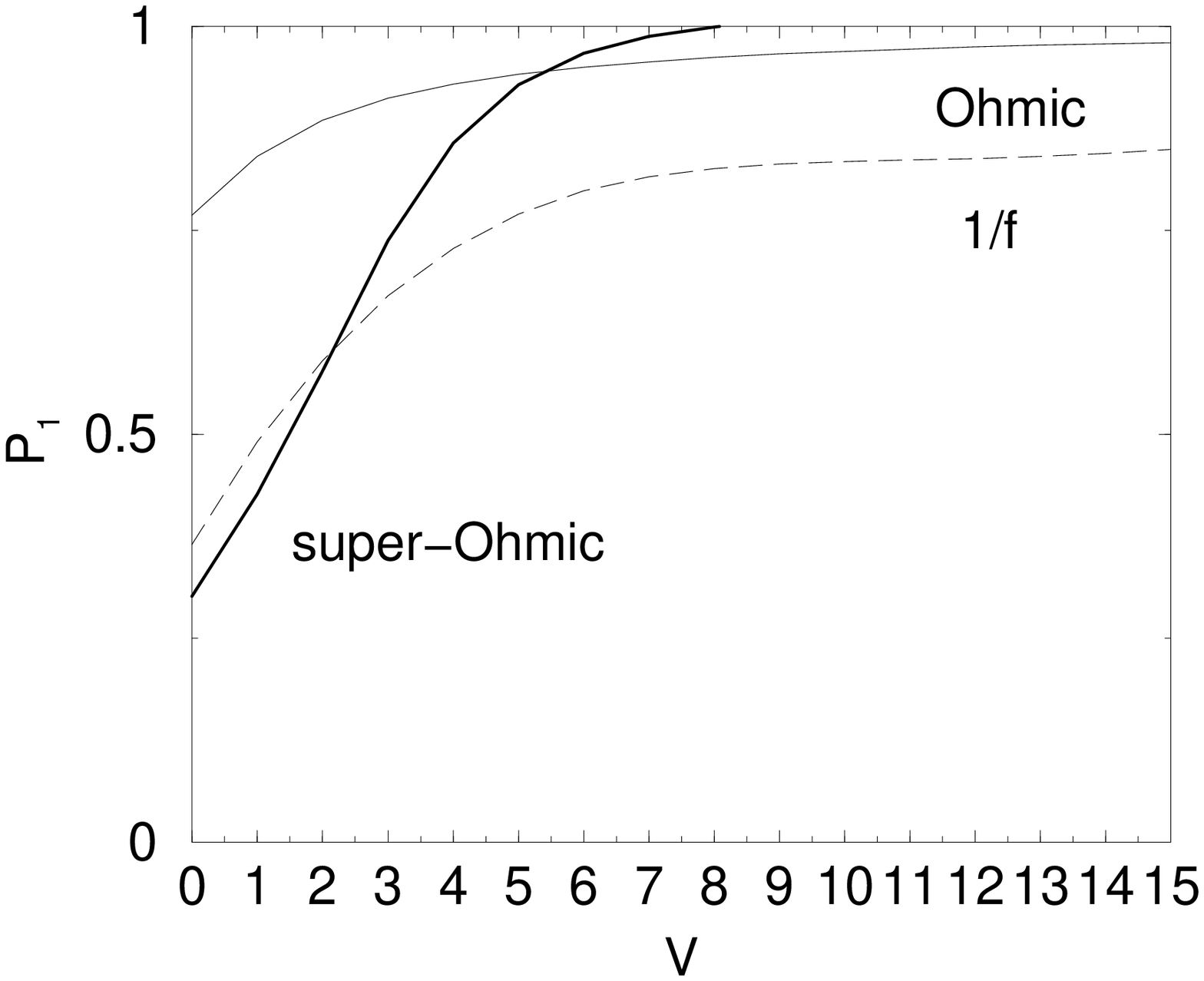}
  \end{center}
\caption{ Excited state population as a function of $V$ is
plotted at $t=0.15[ns]$. The initial state is the first excited
state. $\Lambda_{UV}=100[GHz]$. $\Omega=1[GHz]$. Parameter values
are: $\gamma=0.1[GHz]$ for Ohmic, and $\gamma=0.1[GHz]$ for 1/f,
$\gamma=0.001[GHz]$ for super-Ohmic case.}
\end{figure}

\subsection{Strong, long and slow pulse limit}
We now study the limit of long ($\tau\rightarrow \infty$) and
slow ($\Delta t\rightarrow \infty$) pulses, opposite to the cases
considered so far. In this limit, the pulse Hamiltonian can be
written as the interaction between the system Hamiltonian and the
constant field:
\begin{eqnarray}
H_P\rightarrow H_S V.
    \label{HV}
\end{eqnarray}
Let us consider the time evolution operator acting on the eigenstate of the total Hamiltonian:
\begin{eqnarray}
U(t) |N\rangle =e^{-i E_N t } |N\rangle, \label{Upert}
\end{eqnarray}
where $H |N\rangle = E_N |N\rangle$. We take the large $V$ limit
and consider the first order correction with respect to $V^{-1}$.
Using the nondegenerate perturbation theory, we see that
$|N\rangle =|n\rangle + V^{-1} |n\rangle^{(correction)}$, $E_N =V
\epsilon_n + \langle n | H_S + H_I + H_E |n\rangle +O(V^{-1})$,
where $|n\rangle$ and $\epsilon_n=\Omega n$ are eigenstates and
eigenvalues of the Hamiltonian $H_S$. By expanding (\ref{Upert})
around these unperturbed states, we obtain in the large $V$ limit,
\begin{eqnarray}
U(t) |N\rangle &=& e^{-i E_N t } |n\rangle +O(V^{-1})\nn &=& e^{-i
(V \epsilon_n + \langle n | H_S + H_I + H_E |n\rangle +O(V^{-1}))
t } +O(V^{-1})\nn &=& e^{-i ((V+1) \epsilon_n + H_E ) t }
+O(V^{-1})\nn \label{Upertexp}
\end{eqnarray}
for all $N$. This implies
\begin{eqnarray}
U(t) &=& e^{-i ((V+1) H_S + H_E ) t } +O(V^{-1})\nn \label{Upertexpfinal}
\end{eqnarray}
Thus in the limit of large $V$, the system-bath interaction $H_I$
is eliminated. This effect resembles the quantum Zeno
effect\cite{MisraSudarshan77} in which the decay is suppressed by
frequent measurements. Indeed a parallel argument with the
quantum Zeno effect case is possible (\ref{Upertexpfinal}) and
one can show that there is no relaxation in this
limit\cite{FacchiPascazio02}. From (\ref{Upertexpfinal})
\begin{eqnarray}
\rho_{nn}(t) &=& \langle n |\hat{\rho}(t)|n\rangle \nn &=&
\mbox{Tr} \left[U(t) \hat{\rho}(0) U^{-1}(t) |n\rangle\langle n | \right]\nn
&=& \mbox{Tr} \left[\hat{\rho}(0) |n\rangle\langle n | \right] =P_n(0).
\label{SSR}
\end{eqnarray}
The difference here from the quantum Zeno effect in
\cite{MisraSudarshan77} caused by repeated measurements is that
the time evolution of the system in this case is unitary. The
effect of the external field in (\ref{HV}) is to slow down the decay process.
It can in principle provide an alternative method to dynamical decoupling
although $V$ is often uncontrollable as it is fixed by the design
and fluctuations of $V$ induce an additional source of
decoherence. In Fig. 4, the excited state population is plotted
as a function of the strength of the constant external field $V$.
 For Ohmic and super-Ohmic cases, frequency renormalization (Ohmic)
and frequency and mass renormalization (super-Ohmic) terms are
taken into account by introducing counter
terms\cite{ShiokawaKapral02} before applying $V$. These terms
dominant at ultra-short times are not suppressed by the constant
field in (\ref{HV}) while they can be suppressed by the pulse
decoupling (\ref{HP}) as we saw in Sec.3.2 and 3.3. In Fig. 4, the
decay can be suppressed from coupling to constant large external
field $V$ after the renormalization although overall suppression
is not so remarkable as decoupling by pulses . In pulse
decoupling, the effect of the environmental mode with frequency
$\omega<<1/\Delta t$ is suppressed, while in the constant field
case, for any finite $V$, all the environmental modes upto
$\omega\sim 1/t$ contribute to decoherence.

\subsection{Conclusion}\label{Conclusion}

In this paper, we propose a dynamical decoupling scheme for error
prevention or deterrence based on the effective spin-boson model
we introduced recently. A train of decoupling pulses with
alternating signs creates a fictitious instantaneous Hamiltonian
evolution. Our analysis does not make any of the conventional
approximations such as the Born, Born-Markov, RWA or adiabatic
approximation, thus enabling us to probe the system dynamics in a
much wider parameter regime including the low temperature,
ultra-short time scale and strong external field conditions which
are likely to be more relevant for quantum information processing.
We studied this method in a general environment with Ohmic,
super-Ohmic and $1/f$ spectral density functions. In all cases, in
our decoupling scheme, there is no saturation of decoherence rate
due to resonance as reported in previous studies using bang-bang
controls. We saw the smooth crossover from suppression to
accentuation of decoherence in our model that extends in a wide
range of pulse durations. In particular, our method can suppress
$1/f$ noise with much slower pulses than previously studied.  Due
to the shielding of high energy modes in the environment as a
result of defect motion, decoherence caused by telegraph noise is
weaker at short times than that by $1/f$ noise from the harmonic
environment discussed in this paper. Thus we expect our pulse
decoupling method could also provide efficient suppression of
telegraph type noises.

For efficient suppression of decoherence from an Ohmic
environment, the pulse parameter needs to satisfy $\eta < 1$ and
the pulses need to be fast compared to the UV cutoff, which is
difficult to implement. For 1/f noise, $\eta > 1$, thus the
pulses can be reasonably slow and our scheme is relatively easy to
implement. However, for super-Ohmic environment, $\eta << 1$ and
the pulses need to be ultra-fast. Thus the most serious obstacle
for the implementation of quantum computing may be the presence of
super-Ohmic environments.

We also studied the effect of strong constant field on the
population decay. The decay process appears to slow down but the
suppression is not so evident compared to the fast pulse
decoupling.

In reality, we may not know enough information about the
environment to identify the possible sources of decoherence
\cite{IontrapQtelportation}. But one can cleverly reverse the
argument here and propose to use our decoupling method to obtain
information about the environment spectrum by measuring the
decoherence rate with changing pulse intervals.  For instance, if
the value of $\eta$ at which the crossover between decoherence
suppression and accentuation occurs is greater than $1$, the
environment has sub-Ohmic spectrum. If smaller than $1$,
super-Ohmic.

At this point, we are not aware of any experimental results on
super-Ohmic environments. Since an electromagnetic field seen by
a charged electron due to charge-field interaction in a three
dimensional space is a super-Ohmic
environment\cite{BaroneCaldeira91} we expect this type of
environment to be rather ubiquitous than peculiar. While the $1/f$
noise due to charge-fluctuations in Coulomb forces are frequently
observed and more dominant, super-Ohmic noise due to
charge-photon interaction is likely to coexist with $1/f$ noise.
However, the short time dominant characteristics of super-Ohmic
noise will make the actual detection very difficult.

In quantum computation with multiqubits, the entanglement among
qubits is used as a resource for the speed-up of computation based
on quantum algorithm. In the presence of entangling operation, the
propagation of error in the whole computing space is unavoidable.
To overcome this, fault tolerance is required in quantum error
correction schemes\cite{Shor97}. The spreading of non-Markovian
noise, in particular, noise of the super-Ohmic type in the
computation architecture can be a serious threat to many proposed
fault-tolerant computation schemes \cite{AHHH02}.

Our results are readily applicable to the study of quantum
aspects in nanomechanical resonators\cite{ClelandRoukes96} for
which the system is commonly treated as a harmonic oscillator.
ESBM\cite{ShiokawaHu04} captures the essential short time dynamics
common in many qubit models, in which their computational Hilbert
space is obtained by the lowest two levels of the underlying
multi-level structure. It also allows us to calculate the precise
non-Markovian time evolution of leakage. The leakage is commonly
neglected or crudely estimated in most of the popular theoretical
models. Another application of our work is to superconducting
Josephson junction qubits\cite{JJ,VACJPUED02,CNHM03}, in
particular, the phase qubit models\cite{YHCCQW02}.

%%%%%%%%%%%%%%%%%%%% BL Changes end here %%%%%%%%%%%%%%%%

%%%%%%%%%%%%%%%%%%%%%%%%%%%%%%%%%%%%%%%%%%%%%%%%%%%%%%%%%%%%
\section*{Acknowledgments}
This work is supported in part by a NSF-IRT grant PHY-0426696 and
an ARDA contract MDA90401/C0903.

\appendix
\renewcommand{\theequation}{\thesection\arabic{equation}}
\setcounter{equation}{0}

\newpage

%%%%%%%%%%%%%%%%%%%%%%%%%%%%%%%%%%%%%%%%%%%%%%%%%%%%%%%%%%%%%%%%%%%%%%%%%%%%%%

%%%%%%%%%%%%%%%%%%%%%%%%%%%%%%%%%%%%%%%%%%%%%%%%%%%%%%%%%%%%%%%%%%%%%%%%%

\end{document}